# Experimental Measurement of the Berry Curvature from Anomalous Transport

Martin Wimmer[1,2], Hannah M. Price[3], Iacopo Carusotto[3], Ulf Peschel[2]

[1]Erlangen Graduate School in Advanced Optical Technologies (SAOT), 91058 Erlangen, Germany

[2]Institute of Solid State Theory and Optics, Abbe Center of Photonics, Friedrich Schiller University Jena, Max-Wien-Platz 1, 07743 Jena, Germany

[3]INO-CNR BEC Center and Department of Physics, University of Trento, via Sommarive 14, 38123 Povo, Italy

**Introductory Paragraph:**

**Geometrical properties of energy bands underlie fascinating phenomena in a wide-range of systems, including solid-state materials, ultracold gases and photonics. Most famously, local geometrical characteristics like the Berry curvature can be related to global topological invariants such as those classifying quantum Hall states or topological insulators. Regardless of the band topology, however, any non-zero Berry curvature can have important consequences, such as in the semi-classical evolution of a coherent wave packet. Here, we experimentally demonstrate for the first time that wave packet dynamics can be used to directly map out the Berry curvature. To this end, we use optical pulses in two coupled fibre loops to study the discrete time-evolution of a wave packet in a 1D geometrical "charge" pump, where the Berry curvature leads to an anomalous displacement of the wave packet under pumping. This is both the first direct observation of Berry curvature effects in an optical system, as well as a proof-of-principle demonstration that wave packet dynamics can serve as a high-resolution tool for mapping out geometrical properties.**

**Manuscript:**

The Berry curvature is a geometrical property of an energy band, which plays a key role in many physical phenomena as it encodes how eigenstates evolve as a local function of parameters[1]. In a 2D quantum Hall system, for example, the integral of the Berry curvature over the 2D Brillouin zone (BZ) determines the Chern number: a global topological invariant that underlies the quantization of Hall transport for a 2D filled energy band[2]. As first explained by Thouless[3], there can be an analogous topological quantization of particle transport in a 1D band insulator, when the lattice potential is "pumped" i.e. is slowly and periodically modulated in time. Also in this case, the geometrical and topological properties are defined for an effective 2D parameter space, but now one spanned by the 1D Bloch momentum and the external periodic pumping parameter.

A local non-zero Berry curvature can have striking physical effects in both 2D systems and 1D pumps, regardless of whether the global topological Chern number is non-trivial. In the simplest case, a semi-classical wave packet moves as a coherent object governed by classical equations of motions with an additional "anomalous" Hall velocity due to the geometrical Berry curvature at its center-of-mass, as an external force is applied or as the control parameter is pumped[1,4]. As highlighted further below, this anomalous transport can be understood physically as the Berry curvature acting like a magnetic field in parameter space[5–7].

In recent years, there have been many landmark experiments to engineer and study geometrical and topological energy bands in ultracold gases and photonics[8–28]. In photonics, for example, topological edge states have been studied in a wide-variety of (effectively) 1D set-ups, such as quantum walks[14] and pumping in optical quasicrystals[13,29,30], as well as in 2D quantum Hall-like systems of photonic crystals[9–11], propagating wave-guides[12] and silicon ring resonators[15,16]. However, while the existence of such edge states is linked to the non-trivial global topological properties of bulk bands, a direct measurement of the Berry curvature itself has so far not been realised. In ultracold atoms, instead, there has been much progress in this direction, for example, with experiments exploiting the effects of Berry curvature on particle transport to extract, in 2D systems, the Chern number[19] and topological phase diagrams[20], and, in 1D systems, both the quantized "Thouless pumping" of atoms in a filled band[23,24] and the un-quantized "geometrical pumping" of a Bose-Einstein condensate[25].

In this Letter, we go beyond previous experiments to provide the first direct measurement of the Berry curvature in an optical system. We also demonstrate the first full tomography of the Berry curvature of a geometrical energy band from measurements of the anomalous displacement of a coherent classical wave packet. This approach complements existing methods in ultracold gases to measure the Berry curvature via band-projection[26] or using Wilson lines[27], with the significant advantage that it does not require a full reconstruction of the band eigenstates as an intermediate step. Consequently, our method can be easily applied to complex systems with many bands and degrees of freedom, where a full determination of the eigenstates would be experimentally challenging.

In our "photonic mesh lattice" set-up[31,32], pulses are injected into two coupled optical fiber loops, so that the temporal evolution of the outgoing pulse-train envelope simulates a wave packet in a 1D lattice. More specifically, the position of the particle is encoded in the time at which a light pulse completes a round-trip through one fiber loop. Tunneling between time slots is achieved as the two coupled loops have slightly different lengths, allowing the pulse to "hop" from one time slot to another. We then apply a periodic phase modulation in one loop to engineer Floquet quasi-energy bands, in which the eigenstates depend on both the 1D "Bloch momentum", $Q$, and on a control parameter, $\varphi$, given by the amplitude of the phase modulation. With respective to these Floquet bands, the Berry curvature takes the form[1]

$$\Omega_j^{\varphi,Q} = \frac{\partial}{\partial \varphi}<\psi_j\left|i\frac{\partial}{\partial Q}\right|\psi_j> - \frac{\partial}{\partial Q}<\psi_j\left|i\frac{\partial}{\partial \varphi}\right|\psi_j> \qquad (1)$$

where $|\psi_j>$ are the eigenstates making up the $j$-th band in the effective 2D parameter space. By preparing a wave packet with a narrow spectral width and a well-defined center-of-mass momentum $Q$, we realise a geometrical charge pump by slowly ramping the phase amplitude, $\varphi$, in time. Due to this pumping, the wave packet moves along the 1D "photonic mesh lattice", with an anomalous velocity contribution[1,23–25] due to the Berry curvature $\Omega_j^{\varphi(t),Q}$. Our high-resolution observation of the wave packet dynamics then allows us to map out the Berry curvature for an optical energy band for the first time. By bringing geometrical energy bands into the new setting of photonic mesh lattices, our work opens up new opportunities for engineering novel systems, combining topology, nonlinearity, gain or loss, which may then be studied via wave packet dynamics.

In our experiment, we employ a time-multiplexing, linear optical scheme based on two coupled fiber loops of different lengths[31,32] as shown in Fig. 1a. At the beginning of each measurement, an optical pulse with a wavelength $\lambda = 1555$ nm is injected into the longer loop before being split by a 50/50 coupler into one pulse in each loop. After every round-trip, each pulse is again split into two, although due to the different loop lengths, the short loop pulse completes the round-trip in a time, $T_2$, which is shorter than that taken in the long loop, $T_1$. Crucially, the average loop-length is much larger than the difference in loop lengths; in particular, the former is about 1 km, corresponding to a $\bar{T} = (T_1 + T_2)/2 \approx 5192$ ns average round trip time, while the latter is only about 7 m, corresponding to a time-difference of $\Delta T = T_1 - T_2 \approx 35.4$ ns between round trips in the long and short loops. Furthermore, in order to avoid any overlap between distinct pulses, the pulse width $T_{\text{pulse}}$ is set to 25 ns, such that it is shorter than the minimum pulse-separation $\Delta T$. Thanks to the separation of all of these time-scales, we observe clearly-separated sequences of pulses over time (see Fig. 1), so that each pulse can be straightforwardly labelled by two integers $(m, n)$, with $m$ being the total number of round-trips and $n$ counting how many more round-trips were made in the long compared to the short loop. Of course, this picture is restricted to sequences shorter than the round-trip time in order for the assignment of integers to be unambiguous; this limits the maximum number of pulses in a given sequence to $\bar{T}/\Delta T \approx 147$.

To detect the average intensities of pulses in both loops, we use a combination of a photodiode and a digital sampling oscilloscope with a bandwidth of 200 MHz and a temporal distance of the samples of 0.4 ns. Thanks to the long pulse-width, $T_{\text{pulse}}$, dispersive effects in the optical fibers are typically unimportant, even over 400 roundtrips, corresponding to 400 km of propagation. However, for removing noise and for modulating the amplitude and phase of the pulses, band-pass filters and modulators are needed, as discussed in detail in the Supplementary Information. These additional components introduce losses on the order of approximately 8dB per roundtrip, which are compensated by Erbium doped fiber amplifiers (EDFA). In general, the use of these amplifiers destroys quantum correlations, however, all experiments here are performed with a classical, coherent bright light pulse and do not rely on quantum effects. Besides compensating any losses, the EDFAs also allow for the use of pulses with a peak power of up to 200 mW. In the future, when combined with dispersion-compensating fibers, this system can be used, therefore, for the measurement of nonlinear effects e.g. the formation of solitons based on nonlinear self-phase modulation, which will be of great interest to study in the context of geometrical or topological energy bands. Additionally, this experimental platform also allows for the investigation of the topology of non-Hermitian systems, by including gain and loss in a controlled way. This can be achieved, by using fiber-coupled acousto-optical modulators, which are set to a transmission of 50%. Depending on an electrical RF signal, this value can be dynamically increased or decreased resulting in an effective gain or loss [32].

Based on the labelling of pulses according to $(m, n)$, the evolution of light can be mapped to a 1+1 dimensional lattice as depicted in Figs. 1b,c, where $m$ is the number of discrete time steps on the vertical axis, and $n$ is the position along the horizontal axis, which can either increase or decrease by one at each time step. In our set-up, the evolution between each time-step is:

$$u_n^{m+1} = \frac{1}{\sqrt{2}}(u_{n+1}^m + iv_{n+1}^m)e^{+i\Phi(m)} \qquad (2.1)$$

$$v_n^{m+1} = \frac{1}{\sqrt{2}}(v_{n-1}^m + iu_{n-1}^m), \qquad (2.2)$$

where $u_n^m$ and $v_n^m$ denote the pulse amplitudes incident on the beam-splitter from the short and long loops. As can be seen in these equations, pulse amplitudes at the $m + 1$ time-step are an equal superposition of pulses at the previous time-step $m$, with a relative $\pi/2$ phase-shift for light that couples from one loop to the other, due to the 50/50 beam-splitter coupling the two loops. As mentioned above, we have neglected here dispersion effects and assumed that the pulse shape does not depend on the specific route taken. In the short loop, we have inserted an additional phase modulator that provides a time-dependent phase-shift $\Phi(m)$. A suitable choice of this phase-shift will generate the non-trivial geometrical Berry curvature as discussed in the following and in the Supplementary Information.

Without the phase-shift $\Phi(m)$, the lattice depicted in Fig. 1b and described by the above equations Eqn. (2) is periodic under a double-step in the time-step $m \to m + 2$ and the position $n \to n + 2$. Guided by the tight-binding analogy above, we apply the so-called Floquet-Bloch ansatz for the eigenstates[32,33]

$$\begin{pmatrix} u_n^m \\ v_n^m \end{pmatrix} = \begin{pmatrix} U \\ V \end{pmatrix} e^{-\frac{im\theta}{2}} e^{\frac{inQ}{2}}, \qquad (3)$$

where the parameter $Q$ is the "Bloch momentum" along the 1D lattice, introduced above, and $\theta$ is the propagation constant or "quasi-energy". Here, the periodic part of the eigenstates in the $j$-th band is denoted by $\psi_j = (U, V)^T$ and the quasi-energy band structure satisfies[33]

$$\cos\theta = \frac{1}{2}(\cos Q - 1), \qquad (4)$$

leading to two bands that are periodic in both $Q$ and $\theta$ as expected in a Floquet system, with a band-touching at the edge of the 1D Brillouin zone where $Q = \pm\pi$ (see Supplementary Note S2). In general, the number of bands can be increased by using a "spatially"-structured phase modulation, i.e. a modulation dependent on position $n$, which increases the size of the unit cell[32], or by adding additional fiber loops of different length, which can also increase the effective "spatial" dimensionality.

We can engineer the time-dependent phase-shift $\Phi(m)$ to realize energy bands with nontrivial geometrical properties, by, for example, cyclically applying a positive phase $+\varphi$ for even time steps and $-\varphi$ for odd (see Fig. 2a), and hence breaking a symmetry of the 1D lattice (see Supplementary Material for further discussion). Experimentally, this means switching the sign of the phase modulation every time that all pulses complete another round-trip. Therefore, the necessary modulation frequency of the phase modulator is on the order of $1/\bar{T} \approx 200$ kHz. As the double-step periodicity is maintained, we again use the above ansatz Eq. (3) and find two bands satisfying

$$\cos\theta = \frac{1}{2}(\cos Q - \cos\varphi), \qquad (5)$$

as plotted in Figs. 2b,c (see also Supplementary Information). The manifold of eigenstates is now defined over the Bloch momentum $Q$ and the periodic phase amplitude $\varphi$, allowing us to define the Berry curvature in Eq. (1). As shown in Figure 2d, this Berry curvature is singular at points where the two bands touch but otherwise can take large non-zero positive and negative values across the effective 2D parameter space.

To probe the Berry curvature, we prepare an optical wave packet by exciting a superposition of states in the $j$-th band with a well-defined mean Bloch momentum $Q$ and with a narrow momentum spread $\Delta Q \approx 0.05\pi$. This can be achieved in good approximation by initially generating a chain of pulses with a Gaussian envelope in the short and long loop, and by setting the correct phase modulation between the envelopes in both loops[34]. Then, by slowly increasing the magnitude of the phase modulation after every second time step, we "pump" the wave packet along the band in the effective 2D parameter space. We distinguish therefore between two time scales; firstly, there is the fast time scale $m$, which counts the round trips and over which the sign of the phase flips as

$$\Phi(M(m)) = \begin{cases} -\varphi(M(m)), & \text{odd } m \\ +\varphi(M(m)), & \text{even } m \end{cases} \quad (6)$$

while the phase magnitude $\varphi$, i.e. the pumping parameter, is determined by the slower time scale $M(m)$. Here, we let the phase magnitude increase linearly with $M$ as

$$\varphi(M(m)) = \varphi_0 M(m) \quad (7)$$

and set $M(m) = \lfloor m/2 \rfloor$. For simplicity, hereafter in our notation we drop the explicit dependence on $m$. Note that the fast phase variation in Eq. (6) is directly included in the bands thanks to the ansatz in Eq. (3), while the slow variation in Eq. (7) adiabatically "pumps" the system, provided that the rate $\varphi_0$ is sufficiently small compared to the band-gap.

The center-of-mass of the Gaussian envelope then evolves along the 1D lattice with a total velocity

$$v_j(M) = v_j^G(\varphi(M), Q) + \frac{\partial \varphi(M)}{\partial M} \Omega_j^{\varphi(M), Q} \quad (8)$$

where the first term is the usual group velocity from the band dispersion

$$v_j^G(\varphi(M), Q) = \frac{\partial}{\partial Q} \theta_j(\varphi(M), Q) \quad (9)$$

and the second is the anomalous velocity due to the Berry curvature. This is easiest to understand for a wave-packet in a 2D lattice, in which case $\vec{v}_{\text{anomalous}} = -\dot{\vec{Q}} \times \vec{z} \Omega_j^{Q_x, Q_y}$, with the Berry curvature, $\Omega_j^{Q_x, Q_y} = \frac{\partial}{\partial Q_x} <\psi_j | i \frac{\partial}{\partial Q_y} | \psi_j> - \frac{\partial}{\partial Q_y} <\psi_j | i \frac{\partial}{\partial Q_x} | \psi_j>$, playing the role of a magnetic field in the 2D Brillouin zone, which is defined over the Bloch momenta $Q_x$ and $Q_y$[5-7]. This anomalous velocity is then the momentum-space analogue of the Lorentz force $\vec{F} = e\dot{\vec{r}} \times \vec{z}B$, acting on a charged electron in a real-space magnetic field, $B\vec{z}$, perpendicular to the 2D lattice. In a 1D pump, $Q_y$ is replaced by the periodic pump parameter, in a procedure often known as dimensional reduction[35]. As the pump parameter is externally controlled, the dynamics along one direction are frozen out, leaving only the anomalous term in Eq. 8.

Experimentally, what we measure is the shift in the center-of-mass of the Gaussian envelope along the "position" $n$ as a function of the number of modulation periods $M$

$$n_j^{\varphi_0}(M) = \int_0^M \left[ v_j^G(\varphi(M'), Q) + \varphi_0 \, \Omega_j^{\varphi(M'),Q} \right] dM'. \tag{10}$$

We note that the limit of topological Thouless pumping can be recovered from this expression, if we consider the motion of a filled and gapped energy band after a complete adiabatic pump cycle. Then, the shift is quantized as an integer[3]

$$n_j^{\varphi_0} = \frac{1}{2\pi} \int_{BZ} \int_0^{2\pi} \Omega_j^{\varphi,Q} d\varphi dQ = C_j, \tag{11}$$

where $C_j$ is the topological Chern number, and where we have used that the group velocity integrates to zero over the full 2D parameter space. In our experiment, however, we consider a system in which the energy bands are not gapped and so the assumption of adiabatic evolution across the entire 2D parameter space breaks down and Eq. (11) is not applicable. Rather, we focus here on the local Berry curvature, and, in particular, on ways to exploit the local sensitivity of a spectrally-narrow wave packet to experimentally extract its value.

To cleanly separate Berry curvature from group velocity effects, we apply a variant of the "time-reversal" protocol[36] and compare the propagation (see Fig. 3a) under a reversal of the sign of the phase magnitude (i.e. $+\varphi_0$ and $-\varphi_0$) to find

$$\Delta n_j(M) = n_j^{+\varphi_0}(M) - n_j^{-\varphi_0}(M) = 2\varphi_0 \int_0^M \Omega_j^{\varphi(M'),Q} dM'. \tag{12}$$

where we have used that, for this modulation, $v_j^G(\varphi, Q) = v_j^G(-\varphi, Q)$ and $\Omega_j^{\varphi,Q} = \Omega_j^{-\varphi,Q}$ (see Supplementary Information S4 and S5). By measuring the real space propagation of the wave packets for $\pm\varphi_0$, the dynamical evolution and hence the influence of the non-vanishing Berry curvature is directly recorded. After exciting an initial eigenstate, no further information about the eigenstates is required as we directly measure the Berry curvature from its physical consequences on wave packet dynamics; this is in contrast to other approaches for measuring the Berry curvature[26,27] which require a full reconstruction of the band eigenstates as an intermediate step. The integrated Berry curvature over a complete phase-pumping cycle is shown in Figs. 2e and 3b. Note that this anomalous displacement can become very large as we approach the band-touching points where the Berry curvature diverges and adiabaticity breaks down; experimentally we measure an anomalous displacement of up to approximately 1 position over 200 time steps for a wave-packet centered around $Q = 0.1\pi$ subject to a phase-gradient of $\varphi_0 = \pi/50$. The general applicability of this method is further confirmed in the Supplementary Information S8, where the same experimental protocol is applied to a second example of modulation and pumping sequence.

As the most important advance of our work, the wave-packet evolution can be measured after each discrete time-step, allowing for unprecedented high time-resolution of the dynamics. We exploit this to map out the local Berry curvature directly from anomalous transport for the first time, by performing a discrete differentiation of the lateral shift $\Delta n_j(M)$ as (see Fig. 3c)

$$\Omega_j^{\varphi(M),Q} = \frac{\Delta n_j(M+1) - \Delta n_j(M-1)}{4\varphi_0} \tag{13}$$

The result of this procedure is shown in Fig. 3d,e; as can be seen, the main features of the Berry curvature are well captured, such as the overall symmetry and sign, the quantitative magnitude over much of the band and the predicted complementarity of the two bands[1,26] (i.e. $\Omega_1^{\varphi,Q} = -\Omega_2^{\varphi,Q}$). Due to the "on the fly" tracking of the wavepacket, we can measure the Berry curvature even along trajectories that eventually hit band singularities. Close to the center of the Brillouin zone ($Q \approx 0$), for example, we can extract the Berry curvature until the singularity is hit at $\varphi = \pi$; afterwards, part of the population is transferred to the other band and the measurement is no longer meaningful. For a gapped band, this technique would straightforwardly give the topological Chern number (Eq. (11)) when we integrate the measured Berry curvature over the effective 2D parameter space.

In conclusion, we have demonstrated that anomalous wave packet dynamics can provide an important experimental tool for mapping out the local geometrical properties of energy bands. This versatility of this scheme should allow it to be applied in a variety of configurations, including systems with gain and loss[32] and/or nontrivial topological invariants. In the presence of nonlinearities, this technique could provide a new approach for the creation and manipulation of different wave packet structures, such as solitons. Given its simplicity, our experimental set-up also holds the promise of transferring geometrical and topological concepts into the applied world of optoelectronics.

**Methods**

A detailed description of the experimental platform is included in the Supplementary Material Note 1.

To reduce the noise limit of the data acquisition, the measured propagation was averaged over approximately 25 realizations. For each band, we sequentially prepared wave packets with an initial mean Bloch momentum $Q$, which was scanned from $-\pi$ to $\pi$ in steps of $0.05\pi$. For each value of $Q$, we performed measurements for $-\varphi_0$ and $+\varphi_0$ before then setting a new value of $Q$.

In order to start with a spatially broad and thus spectrally narrow distribution at a specific Bloch momentum $Q$, a single seed pulse is injected into the fiber loop system. By switching off one of the two loops in a cyclic fashion after every second time step $m$, the system is converted into a lossy mesh lattice as discussed in [37]. The resulting propagation is given by a diffusion equation leading to a coherent splitting of the initial pulse into a chain of pulses with a Gaussian envelope. Due to this diffusive motion, the width of the resulting wavepacket grows $\propto \sqrt{m}$. Thus, for 200 time steps a wavepacket with a width of $\approx 8$ postions with respect to the FWHM of the intensity is generated, which provides a sufficiently small spectral width $\Delta Q \approx 0.05\pi$ to map out the local Berry curvature.

To excite a specific Bloch momentum $Q$, we apply a phase of $\varphi = Q$ during the diffusive motion and a phase of $\varphi = (Q \pm \pi)/2$ for the last time step before the actual propagation. The choice of the sign allows for a selection of either the upper or the lower band. As discussed in [34,37], in this way, the appropriate eigenstate corresponding to a Bloch momentum $Q$ is excited. After the preparation of the Gaussian beam over the first 200 time steps, there is no further switching-off of either loop and, instead, we measure the wave packet dynamics during the next 200 time steps, using the phase modulator to adiabatically "pump" the system as discussed in the main text.

The center-of-masses of the wave packets were extracted by fitting Gaussian distributions to the cross-sections for each time step. A detailed explanation of this fitting procedure, together with examples, is included in Supplementary Material Note 6.

The errorbars depicted in Fig. 2e are estimated by the spectral width $\Delta Q \approx 0.05\pi$ of the initial excitation in horizontal direction and by the uncertainty of the fit procedure for a 95% confidence level in the vertical direction.

For the full reconstruction of the Berry curvature, it is necessary to differentiate the measured integrated Berry curvature with respect to the time step $m$ as discussed in Eq. (13). Directly before evaluating the derivative, a moving average filter is applied to the data to decrease the noise level. A span of 5 values was chosen for the averaging. Raw data without averaging are provided in the Supplementary Material.

**Acknowledgements**

M.W. acknowledges financial support from the Erlangen Graduate School of Advanced Optical Technologies. Additionally, M.W. would like to thank Mark Kremer and Arstan Bisianov for fruitful discussions. Furthermore, this project was supported by PE 523/14-1 and by the GRK2101 funded by the DFG. H.M.P was supported by the EC through the H2020 Marie Sklodowska-Curie Action, Individual Fellowship Grant No. 656093 SynOptic. I.C. was funded by the EU-FET Proactive grant AQuS, Project No. 640800, and by Provincia Autonoma di Trento, partially through the project "On silicon chip quantum optics for quantum computing and secure communications (SiQuro)".

**Additional information**

Supplementary information is available in the online version of the paper. Reprints and permissions information is available at [www.nature.com/reprints](www.nature.com/reprints). Correspondence and requests for materials should be addressed to U.P. (ulf.peschel@uni-jena.de).

**Author contributions**

M.W. performed the experiments, all authors contributed to the theoretical background and the interpretation of the measurement.

**Competing financial interests**

The authors declare no competing financial interests.

**Data availability**

The data that support the plots within this paper and other findings of this study are available from the corresponding author upon reasonable request

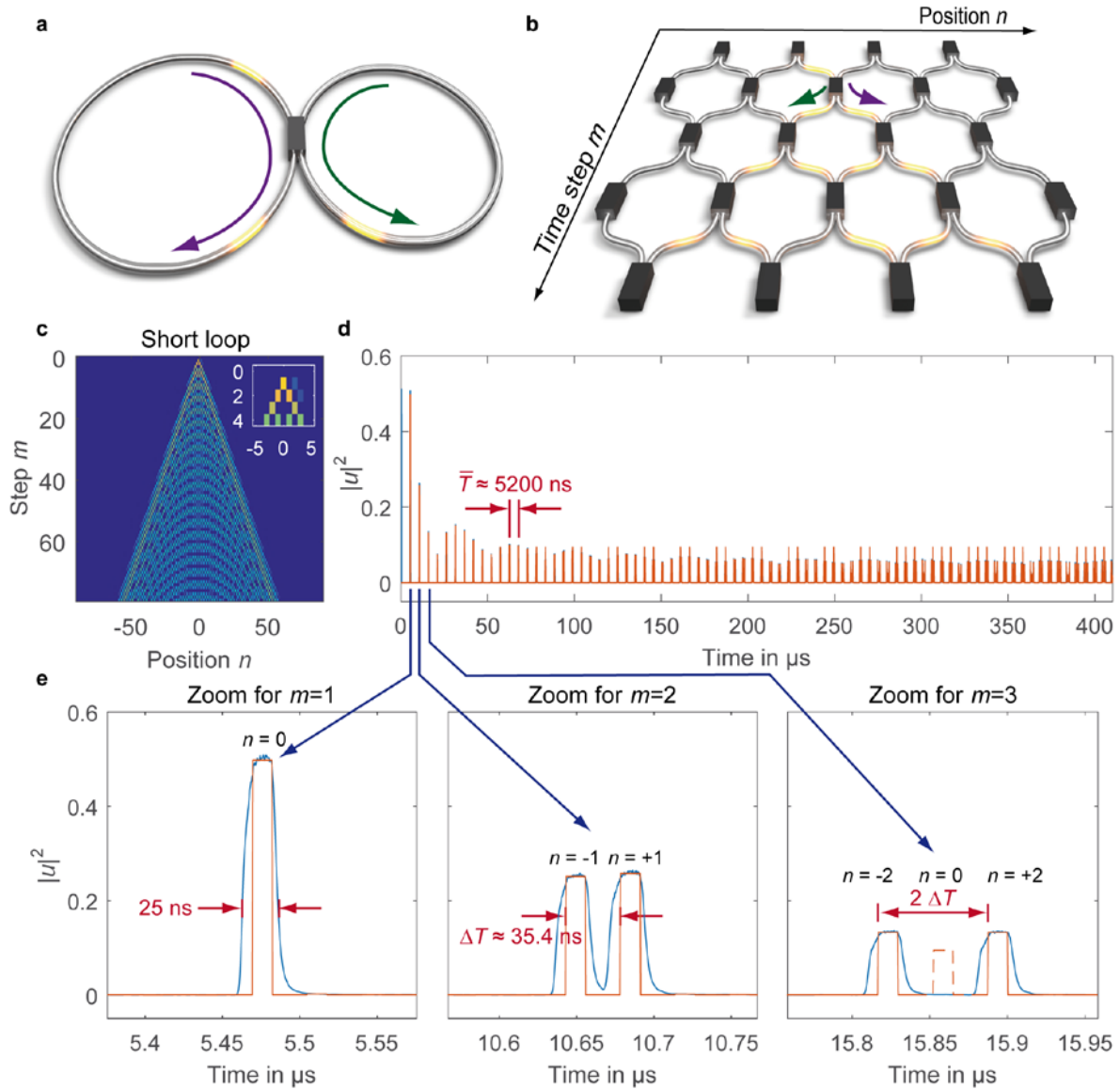

**Figure 1| Generation of photonic mesh lattices by time multiplexing. a,** Optical pulses propagating along two coupled fibre loops of slightly different lengths are used to simulate light evolution on a 1+1D lattice spanned by a discrete time $m$ and position $n$. **b,** The light evolution in the fibre loops is mapped onto a lattice, where a roundtrip in the short loop stands for the motion from northeast to southwest, while the propagation from northwest to southeast is equivalent to a roundtrip through the long loop. **c,** Measurement of the intensity distribution of short-loop pulses for a single lattice site excitation in the absence of any modulation. **d,e,** In the experiment, two photodiodes continuously measure the intensity distribution in both loops; here we show the output from the short loop. We observe a sequence of pulses after every roundtrip time $\bar{T} \approx 5200$ns. Each pulse has a width of $T_{\text{pulse}} \approx 25$ns, and is labelled by its "position" $n$, as shown in (**e**); adjacent pulses within the same sequence are separated by $\Delta T \approx 35.4$ ns, corresponding to the time-difference between an extra round-trip in the long versus the short loop. Here, the blue lines correspond to the signal of the photodiodes and the orange curve to the average height of the pulses. This average height is the quantity shown in panel **c**. For $m = 3$ the first interferences occur, leading to a destructive interference at $n = 0$ (see also panel **c** for $m = 3$).

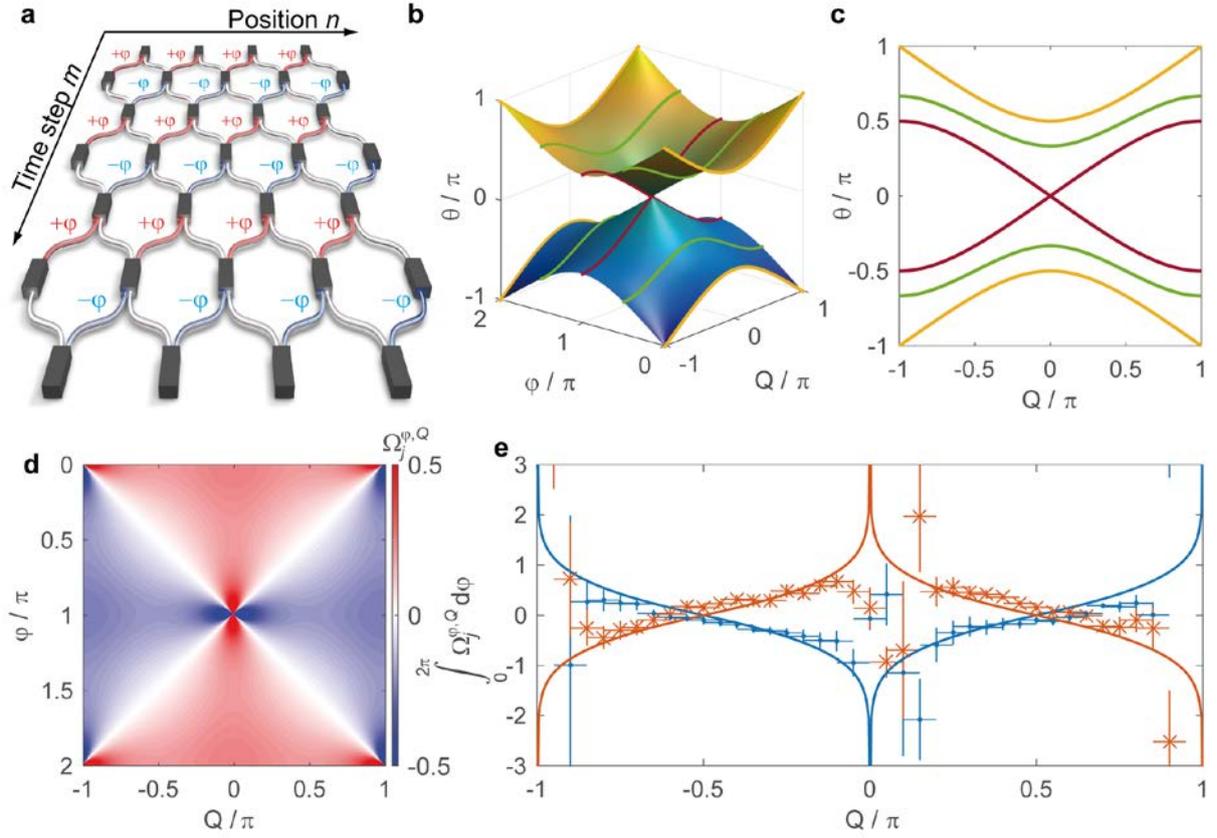

**Figure 2| Anomalous transport of a wave packet. a,** For implementing the geometrical pumping protocol, we apply an additional phase modulation in the short loop with alternating sign at each time step to engineer nontrivial geometrical energy bands. **b,** Band structure of the system as a function of the phase modulation $\varphi$ and Bloch momentum $Q$. Due to the temporal periodicity of the system after two round trips, the band structure is also periodic in $\theta$. **c,** Cut through the band structure for $\varphi = 0$ (yellow), $\varphi = \pi/2$ (green) and $\varphi = \pi$ (red). **d,** Calculated Berry curvature for the lower band ($\theta < 0$); this is singular at the isolated points where the two bands touch at the center and at the corners of the Brillouin zone. **e,** By applying Eq. (12), we extract the integrated Berry curvature over each full experimental trajectory from $\varphi = 0$ to $\varphi = 2\pi$ (the results for the lower band are marked in orange and for the upper band in blue). For Bloch momenta in the vicinity of $Q = 0, \pm\pi$, the trajectory approaches a band-touching point, and so adiabaticity breaks down and the protocol fails. The horizontal errorbars are defined by the spectral width of the wavepacket $\Delta Q \approx 0.05\pi$. The center-of-mass shift used in Eq. (12) is calculated from fitting Gaussian distributions to the intensity distribution after the last step of the evolutions for $\pm\varphi_0$. The uncertainties of these fitting results with respect to the positions of the distributions, are estimated for a 95% confidence interval, squared and summed to define the squared value of the vertical errorbars.

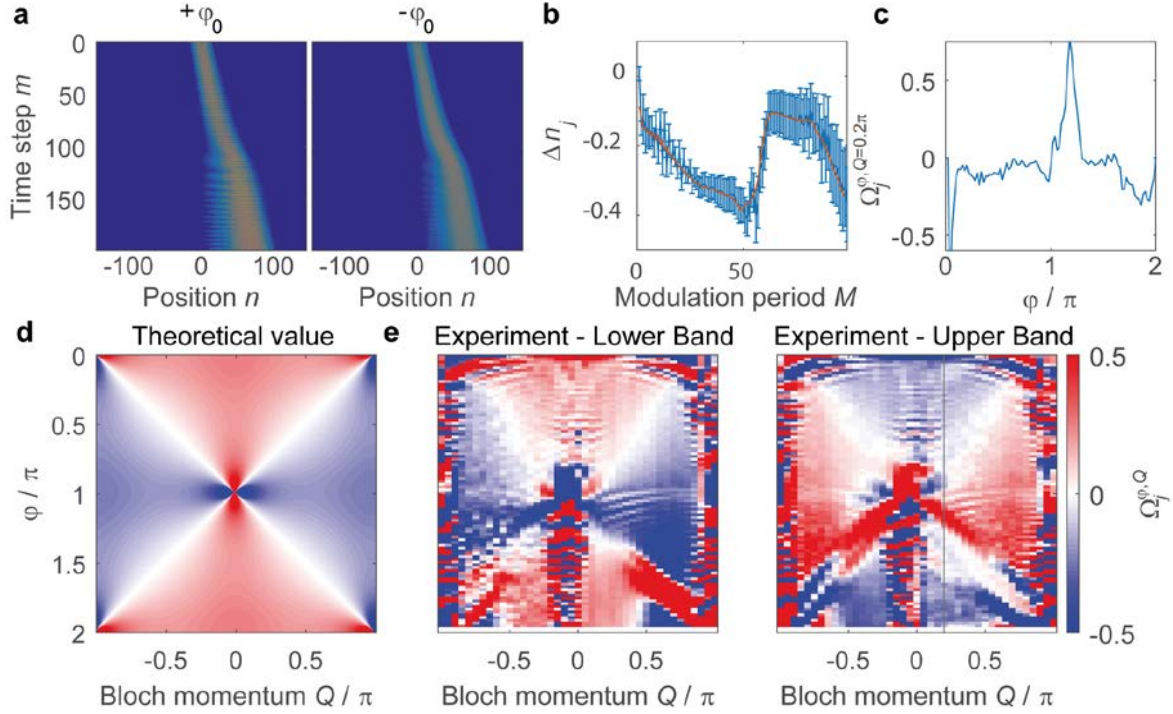

**Figure 3| Experimental reconstruction of the Berry curvature. a,** Measured propagation of wave packets populating the upper band for $Q = 0.2\pi$ for $\pm\varphi_0$. **b,** The relative shift in the centre-of-mass between the two wave packets in panel **a**. Each centre-of-mass is extracted from applying Gaussian fits for each time step $m$. The errorbars are defined according to $\sqrt{\left(\Delta n_j^{+\varphi_0}\right)^2 + \left(\Delta n_j^{-\varphi_0}\right)^2}$, where $\Delta n_j^{\pm\varphi_0}$ are the uncertainties given by the Gaussian fit for trajectories under $\pm\varphi_0$ within a 95% confidence interval. Before numerically differentiating the measured curve (blue line) to find the Berry curvature, a moving average filter with a span of 5 time steps is applied (orange line). **c,** A cut through the Berry curvature for $Q = 0.2\pi$ as a function of the phase amplitude $\varphi$, as obtained by numerically-differentiating the averaged signal in panel **b**. **d,** Calculated Berry curvature for the lower band ($\theta < 0$). **e,** The experimental reconstruction of the Berry curvature for both bands. The cut shown in panel **c** is indicated in the upper band panel by the grey solid line. Compared to the theoretical value for the lower band in (**d**), the experimental results in (**e**) are in a very good agreement and also illustrate the relation of the Berry curvature $\Omega_1^{\varphi,Q} = -\Omega_2^{\varphi,Q}$ for both bands.

# Supplementary Material for Experimental Reconstruction of the Berry Curvature based on the Measurement of Anomalous Transport

Martin Wimmer[1,2], Hannah M. Price[3], Iacopo Carusotto[3], Ulf Peschel[2]

[1]Erlangen Graduate School in Advanced Optical Technologies (SAOT), 91058 Erlangen, Germany

[2]Institute of Solid State Theory and Optics, Abbe Center of Photonics, Friedrich Schiller University Jena, Max-Wien-Platz 1, 07743 Jena, Germany

[3]INO-CNR BEC Center and Department of Physics, University of Trento, via Sommarive 14, 38123 Povo, Italy

## Content

## S1 Experimental setup

A linear version of the coupled fiber loop system presented in [S1] is used. For the signal generation (see Supplementary Fig. S1), a DFB (distributed feedback) laser diode operating at $\lambda = 1555$ nm is cut into a chain of 25 ns long pulses by a Mach-Zehnder modulator (MZM). Afterwards, the pulses are amplified by two successive Erbium-doped fiber amplifiers (EDFA), before another MZM is set to zero transmission between two adjacent pulses in order to ensure a high suppression.

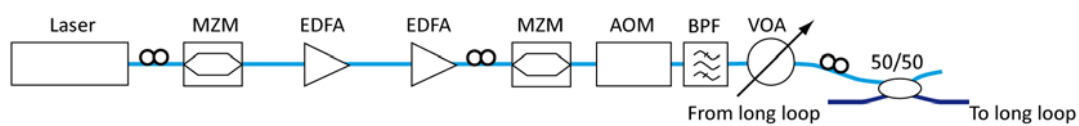

**Supplementary Figure S1 | Signal generation.** The signal of a DFB (distributed feedback) laser is cut into a chain of 25 ns pulses by a Mach-Zehnder modulator (MZM), which is then amplified by two Erbium-doped fiber amplifiers (EDFA). A second MZM is used to suppress the background of the amplified pulse chain, which is adjusted in amplitude by a variable

optical attenuator before the pulses are inserted into the fiber loops. An acousto-optical modulator (AOM) is set to full transmission during the warm-up phase and is set to total absorption after the last warm-up pulse, which is then the seed pulse for the measurement.

In the last stage, the pulses propagate through an acousto-optical modulator (AOM) and are cleaned by a bandpass filter before being injected into the longer fiber loop. At the beginning of the experiment, the AOM is set to a transmission of 100% allowing the pulse chain to enter the fiber loop system. The EDFAs in both loops are sensitive to the input power level on a time scale of approximately 10µs. In order to avoid the excitations of transients, a quasi continuous chain of pulses with a total length of about 8ms is injected into the system. After about 8 ms the system and especially the EDFAs are adapted to the input power level. When the last pulse of the warm-up phase has entered the system, the AOM in the signal generation line (see Supplementary Fig. S1), which is called entrance AOM, is set to 0% transmission and thus blocking all succeeding pulses from the signal generation.

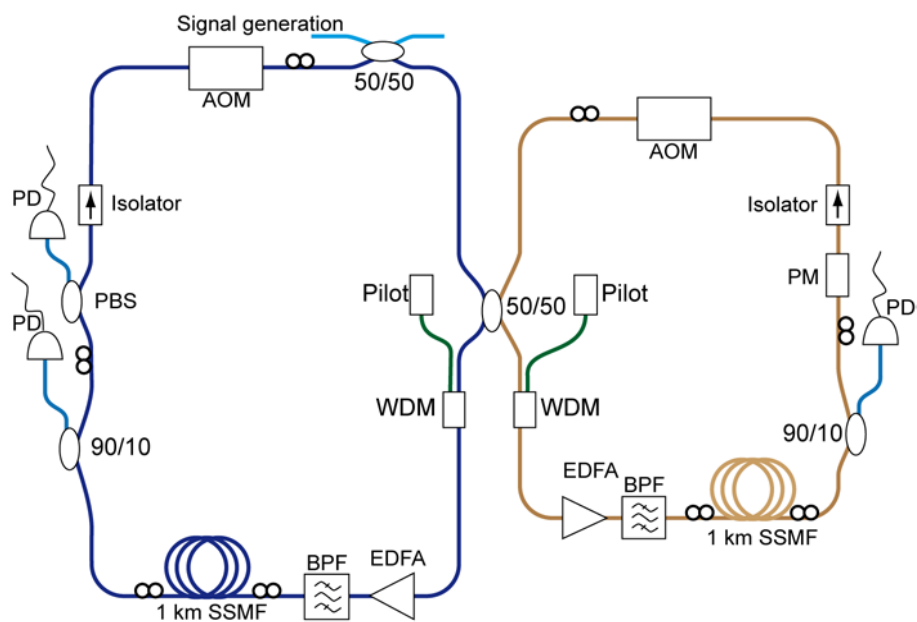

**Supplementary Figure S2 | Sketch of the experimental fiber loop setup.** Both fiber loops are connected by the 50/50 coupler. Any losses during one round trip are compensated in each ring by an Erbium-doped fiber amplifier (EDFA). In order to maintain a constant amplification rate, a pilot signal is inserted via a wavelength division multiplexing coupler (WDM) and afterwards removed by a bandpass filter (BPF). In order to establish a long average round trip time, each loop contains a 1 km long fiber spool followed by a 90/10 monitor coupler and a photo diode (PD). In the long loop, a polarizing beam splitter (PBS) is used to select a single polarization, while in the short loop a phase modulator (PM) allows for the generation of an arbitrary potential. An isolator prevents any back reflected light from circulating through the system. Acousto-optical modulators (AOM) at the end of each fiber ring are needed to block the circulation of pulses during the warm up phase.

Inside the coupled fiber loop system (see Supplementary Fig. S2), the seed pulse first splits up at the 50/50 coupler connecting the longer and shorter loop. In each fiber ring, a wavelength division multiplexing coupler (WDM) is used to add a blue-shifted strong continuous wave (CW) pilot signal operating at about 1537 nm into the fiber ring. The pilot signal maintains a constant amplification factor of the EDFAs in each loop and is removed from the signal by bandpass filters directly after the amplifiers. The pulses propagate through standard single mode fiber (SSMF) spools with a length of about 1 km. In order to fine tune the length difference, SSMF patch cords are used. In total the average round trip time of both fiber loops is about 5200 ns while the length different is chosen to be approximately 35 ns. The number of observable positions is equal to the ratio of the two time constants $n_{max} = \frac{5200}{35} \approx 150$. For monitoring the pulses circulating in the system a photodiode (PD) is attached to each loop via a 90/10 coupler. Furthermore, a single polarization is selected by a polarizing beam splitter in the longer loop. In the short loop, a phase modulator (PM) is used to modulate the pulses according to Eq. (6) of the main paper and Eq. (S4) of the Supplementary Material. Finally, an AOM is placed in each fiber loop, which is set to zero transmission during the warm up phase, so that no pulses can recirculate through the fiber loops. When the AOM in the signal generation is set to no transmission, the loop AOMs are on the contrary set to full transmission and allow for the circulation of the initial seed pulse. Furthermore, the loop AOMs are used for the generation of Gaussian pulse trains as is explained in the Supplementary Material of [S1].

## S2 Floquet band structure

Without any phase modulation, the size of the unit cell of the lattice depicted in Fig. 1a of the main part covers two rows and positions[S1,S2]. For such a double step, the evolution is described by

$$2u_n^{m+2} = u_{n+2}^m + iv_{n+2}^m + iv_n^m - u_n^m \tag{S1.1}$$

$$2v_n^{m+2} = v_{n-2}^m + iu_{n-2}^m + iu_n^m - v_n^m. \tag{S1.2}$$

Since the system is evolving in discrete time steps, denoted by the superscript $m$, on a discrete lattice, with sites labelled by $n$, we apply a Floquet Bloch ansatz of the form [S2]

$$\begin{pmatrix} u_n^m \\ v_n^m \end{pmatrix} = \begin{pmatrix} U \\ V \end{pmatrix} e^{-\frac{i\theta m}{2}} e^{\frac{iQn}{2}} \tag{S2}$$

The resulting dispersion relation (see Supplementary Fig. S3)

$$\cos\theta = \frac{1}{2}(\cos Q - 1) \tag{S3}$$

consists of two bands separated by a band gap. However, due to the Floquet nature of the system, the band structure is not only periodic in $Q$ but also in $\theta$ and as a result both bands intersect for $Q = \pm\pi$ and $\theta = \pm\pi$ (see Supplementary Fig. 3). The two component vector $(U, V)^T$ stands for the eigenstate corresponding to a given value of $\theta$ and $Q$, and which therefore is determined by the phase- and amplitude-relation between the short and long loop. In good approximation, an eigenstate is excited by generating a Gaussian beam in both loops and by setting the appropriate phase relation[S1]. According to the size of the unit cell, the natural time scale along the propagation direction covers two time steps.

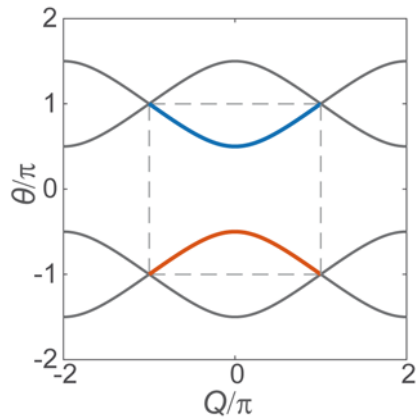

**Supplementary Figure S3 | Dispersion relation of the unmodulated system.** Due to the temporal periodicity of the system after two round trips, the band structure is periodic in the Bloch momentum $Q$ and the propagation constant $\theta$. The first Brillouin zone is highlighted by the dashed gray rectangle.

## S3 Properties of the modulated system

In order to implement the geometrical energy bands, a phase modulation $\Phi$ is applied to the short loop according to

$$\Phi(M(m)) = \begin{cases} -\varphi, & \text{odd } m \\ +\varphi, & \text{even } m \end{cases} \tag{S4}$$

where the sign alternates after each single time step $m$. From the evolution equations (Eq. 2.1 and 2.2 in the main text), we can derive the Hamiltonian of the 1D system for a constant modulation amplitude $\varphi$; as this is a two-band model, the Hamiltonian can be expressed in the form $H = \vec{n} \cdot \vec{\sigma}$, where $\vec{\sigma}$ is the vector of Pauli matrices and here

$$\vec{n} = \frac{\cos^{-1}\left(\frac{1}{2}(\cos Q - \cos \varphi)\right)}{\sqrt{1 - \frac{1}{4}(\cos Q - \cos \varphi)^2}} \begin{pmatrix} \cos Q + \cos \varphi \\ -\sin Q + \sin \varphi \\ \sin Q + \sin \varphi \end{pmatrix}$$

The corresponding dispersion relation is then

$$\cos \theta = \frac{1}{2}(\cos Q - \cos \varphi). \tag{S5}$$

The band structure is depicted in Supplementary Fig. S4 for different values of $\varphi$.

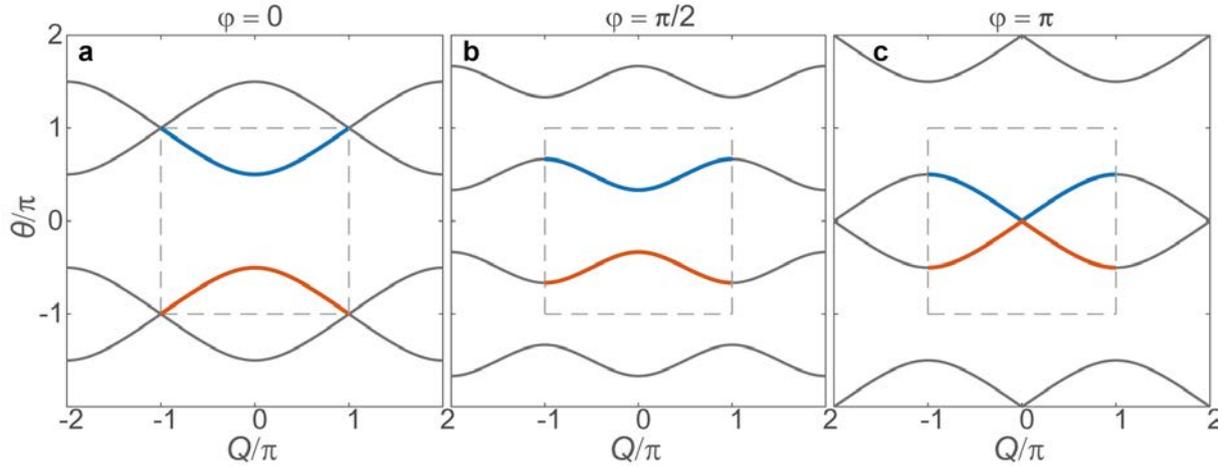

**Supplementary Figure S4 | Band structure of the temporal driven system. a-c,** Band structure of the system driven by the two step modulation for $\varphi = 0$, $\varphi = \pi/2$ and $\varphi = \pi$. For $\varphi = \pi$ band gap is closed at the center of the Brillouin zone.

To induce a non-zero Berry curvature in the effective 2D parameter space spanned by Q and φ, we have had to break a suitable symmetry of the 1D Hamiltonian via our choice of phase modulation. To understand this, we firstly define a time-reversal operator for the 1D lattice and see that it commutes with the Hamiltonian; for our system, we have $\hat{T} = e^{i\frac{\pi}{4}\sigma_x}\hat{K}$, where $\hat{K}$ is the complex conjugation operator. Following Ref. (S5), it is then straightforward to show that the presence of this time-reversal symmetry implies that the Berry curvature is even with respect to the Bloch momentum, i.e.

$\Omega_j^{\varphi,Q} = \Omega_j^{\varphi,-Q}$. Note that, as the phase amplitude is an external parameter rather than a variable, this symmetry differs from that in a 2D system, where time-reversal symmetry forces the Berry curvature to be odd with respect to a reversal of both Bloch momenta[S4] $\Omega_j^{Q_x,Q_y} = -\Omega_j^{-Q_x,-Q_y}$. Secondly, we can define the 1D inversion symmetry in our system as $\hat{P} = \hat{\Pi}\sigma_x$, where $\hat{\Pi}$ corresponds to spatial inversion with respect to the "position" in the 1D photonic mesh lattice. Using a similar argument to that in Ref. (S5), the presence of this symmetry would imply that the Berry curvature must be odd with respect to the Bloch momentum, i.e. $\Omega_j^{\varphi,Q} = -\Omega_j^{\varphi,-Q}$. (Again, this is different to a 2D system, where 2D inversion symmetry implies [S4] $\Omega_j^{Q_x,Q_y} = \Omega_j^{-Q_x,-Q_y}$.) In the presence of both symmetries the Berry curvature must vanish. In fact inversion symmetry is present in our system in case of vanishing phase modulation (φ=0) only. We therefore have to choose the phase modulation to break inversion symmetry. As we still have time-reversal symmetry, the now nonvanishing Berry curvature is indeed even with respect to $Q$, as shown in Figure 3 in the main text.

We note that inversion symmetry would also have been broken by a simpler phase modulation $\Phi(M(m)) = \varphi(M(m))$ (i.e. without flipping the sign of the phase); however, this introduces an additional symmetry by which the eigenstates only depend on the combination $Q + \varphi$, rather than on the two parameters separately. As a result, it is straightforward to show that the Berry curvature again vanishes across the Brillouin zone.

Finally, we consider pumping the system by increasing the phase amplitude slowly in time. In particular, the height of the phase modulation

$$\varphi(M(m)) = \varphi_0 M(m) \tag{S6}$$

is increased after every modulation period $M(m) = \lfloor m/2 \rfloor$, as depicted in Supplementary Fig. S5 for a phase gradient $\varphi_0 = \pi/16$. Here, the comparatively large gradient is only chosen to make the scheme more clear; in the experiment, a much smaller phase gradient of $\varphi_0 = \pi/50$ is used to ensure that the wave packet moves adiabatically for most values of the Bloch momentum $Q$. This fails close to the band-degeneracies where adiabaticity breaks down and there can be transitions between bands. Other limitations in the applicability of our protocol to measure the Berry curvature are discussed in Supplementary Note S5.

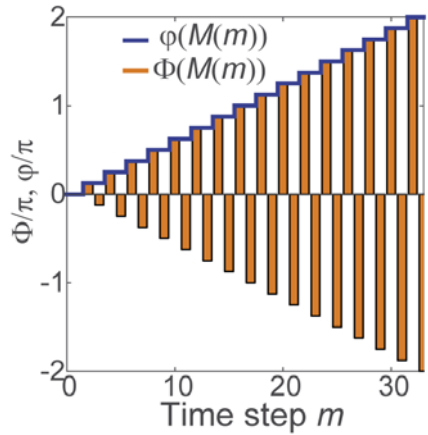

**Supplementary Figure S5 | Illustration of the phase modulation for a gradient $\varphi_0 = \pi/16$.** The applied phase $\Phi(m)$ alternates between $\pm\varphi(m)$ after each roundtrip (see the orange bars), while the amplitude of the phase modulation $\varphi(m)$ increases only after every second time step (see the blue stair function).

## S4 Explanation of the experimental protocol

We consider the dynamics of a wave packet, prepared in a single band $j$, with a well-defined center-of-mass position $n$ and center-of-mass momentum $Q$. In order to understand the dynamical evolution of the system, we first review the case for the modulated system described by Eq. (S4), where the phase amplitude $\varphi = $ const. Then, the total velocity at which the wave packet travels is given by the group velocity of the band

$$v_j^G(\varphi, Q) = \frac{\partial}{\partial Q} \theta_j(\varphi, Q). \tag{S7}$$

After $M$ periods, the center-of-mass position of the wave packet is given by the integral over the group velocity

$$n_j(M) = \int_0^M v_j^G(\varphi, Q) dM' = v_j^G(\varphi, Q) M. \tag{S8}$$

Now, we consider the modulation protocol described by Eq. (S6) where the phase amplitude is gradually ramped up. Then, not only does the group velocity change according to the phase $\varphi(M(m))$, but the wave packet also gains an anomalous velocity due to the Berry curvature, leading to a total velocity[S3,S4] of

$$v_j(Q, \varphi) = v_j^G(\varphi(M), Q) + \varphi_0 \, \Omega_j^{\varphi(M), Q}, \tag{S9}$$

where hereafter we drop the explicit dependence on $m$ for simplicity of notation. The Berry curvature is calculated numerically by evaluating the expression

$$\Omega_j^{\varphi(M), Q} = i \frac{\partial}{\partial \varphi} <\psi_j \left| i \frac{\partial}{\partial Q} \right| \psi_j> - i \frac{\partial}{\partial Q} <\psi_j \left| i \frac{\partial}{\partial \varphi} \right| \psi_j>, \tag{S10}$$

where $\psi_j = (U_j, V_j)^T$ is the eigenstate of a band, as defined in the ansatz of Eq. (S2).

In the experiment, we excite a wave packet centered around a specific Bloch momentum by using the modulation scheme described in the Supplementary Material Note 1.6 of [S1]. Then the wave packet propagates through the lattice under an external phase gradient of either $+\varphi_0$ or $-\varphi_0$. For both cases, the center-of-mass $n_j(M)$ is evaluated for integer multiples of the modulation period $M$, so that we can extract the difference in the center-of-mass shift:

$$n_j^{+\varphi_0}(M) - n_j^{-\varphi_0}(M) = \int_0^M \left( v_j(Q, +\varphi_0 M') - v_j(Q, -\varphi_0 M') \right) dM' \tag{S11}$$

between a positive and negative phase gradient $\varphi_0$.

Evaluating Eqn. (S10) and (S11) leads to the final expression

$$n_j^{+\varphi_0}(M) - n_j^{-\varphi_0}(M) = \int_0^M 2\varphi_0 \Omega_j^{\varphi(M'), Q} dM'. \tag{S12}$$

where we have used that $v_j^G(\varphi, Q) = v_j^G(-\varphi, Q)$ and $\Omega_j^{\varphi,Q} = \Omega_j^{-\varphi,Q}$ (see Section S5). Therefore, the integral of the Berry curvature is given by the lateral shift of the measured center-of-masses. To reconstruct the total Berry curvature, the center-of-mass shift is numerically differentiated in order to find

$$\Omega_j^{\varphi(M),Q} = \frac{n_j^{+\varphi_0}(M+1) - n_j^{+\varphi_0}(M-1) - (n_j^{-\varphi_0}(M+1) - n_j^{-\varphi_0}(M-1))}{4\varphi_0}. \tag{S13}$$

## S5 Restrictions on the protocol

In general, there are two conditions on the underlying model, which need to be fulfilled in order to apply the protocol described above. For the derivation of Eq. (S13) it was assumed that the group velocity and the Berry curvature are symmetric in $\varphi$

$$v_j^G(\varphi, Q) = v_j^G(-\varphi, Q) \text{ and } \Omega_j^{\varphi,Q} = \Omega_j^{-\varphi,Q}. \tag{S14}$$

For our modulation scheme, these conditions are fulfilled as a change in the sign of $\varphi$ is equivalent to a shift of the modulation by one time step (see Supplementary Fig. 6). The band dispersion and Berry curvature, however, are gauge-invariant under this transformation and so Eq. (S14) follows. In the case of a modulation scheme where these conditions are not satisfied, a more general protocol according to [S3] should be chosen, where the wave packet velocity is compared at the same value of $(\varphi, Q)$, under positive and negative phase gradients.

Whenever our scheme [Eq. (S12)] is applicable, it has the key advantage over the general protocol [S3] that the extracted Berry curvature is less affected by band degeneracies and the corresponding breakdowns in adiabaticity. To see this, consider a band which is degenerate with a second band at some parameters $(\varphi_1, Q_1)$, and hence, by the symmetry implied by Eq. S14, also at parameters $(-\varphi_1, Q_1)$. In our scheme, we compare the propagation of wave packets after a given number of modulation steps, under opposite phase gradients [Eq. (S12)]. This approach works well until the wave packets reach the band degeneracies where the evolution is no longer adiabatic; for both negative and positive phase gradients, this occurs after $M_1 = \varphi_1/\varphi_0$ modulation periods. In the more general protocol, we compare the motion of the wave packets at the same value of $(\varphi, Q)$, corresponding to $M^{+\varphi_0} = \varphi/\varphi_0$ modulation periods for the positive phase gradient and $M^{-\varphi_0} = (2\pi - \varphi)/\varphi_0$ modulation periods for the negative phase gradient. As always either $M^{+\varphi_0} > M_1$ or $M^{-\varphi_0} > M_1$, this method breaks down for any Bloch momentum, $Q_1$, for which a band degeneracy occurs somewhere in the 2D parameter space.

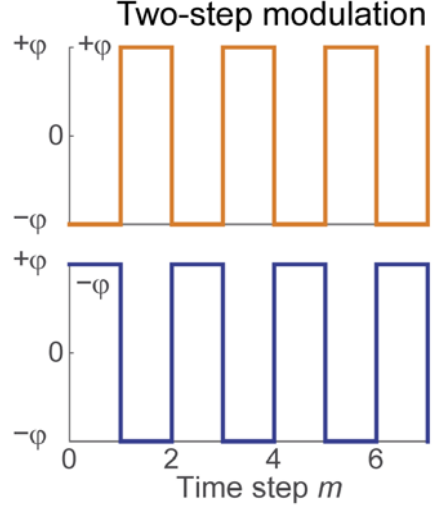

**Supplementary Figure S6 | Comparison of the modulation for $\pm\varphi$ without any phase gradient.** The change in sign of $\varphi_0 \to -\varphi_0$ is equivalent to a shift of the modulation by one time step $m \to m+1$.

## S6 Evaluation of the experimental data

In order to extract the center-of-mass during the propagation, a Gaussian distribution was fitted to each measurement for every second time step, for each Bloch momentum $Q$ (see Supplementary Fig. S7). In some cases, not only one band is excited, but also the other, leading to a secondary wave packet propagating in the opposite direction (see Supplementary Fig. S8c,h) due to the opposite group velocity and Berry curvature in the upper band. The reason for this imperfection can be due to a mismatch of the initial state with the required eigenstate and to the adjacency of band-degeneracy points.

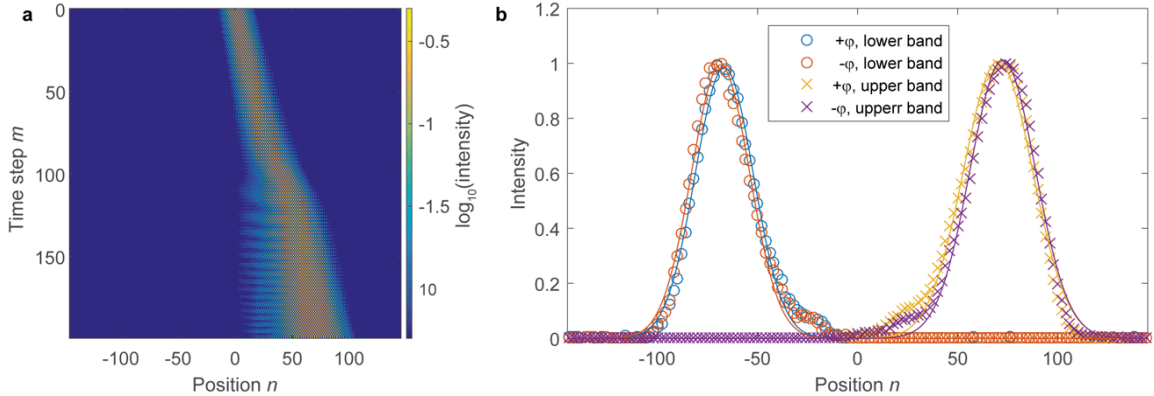

**Supplementary Figure S7 | Example of the propagation of a wave packet and extraction of the center-of-mass. a,** Wave packet propagation for $Q = 0.2\pi$ in the upper band for $\varphi_0 = +\pi/50$. **b,** Extracted experimental cross section, for the last time step $m = 200$, for the upper and lower band and for $\pm\varphi_0$. In order to extract the center-of-mass for each measurement, Gaussian fits (solid lines) are applied to the experimental data points (circles).

For extracting the center-of-mass, a Gaussian distribution of the form

$$f(n) = Ae^{-\frac{(n-n_0)^2}{c^2}} \tag{S15}$$

is fitted to the intensity distribution for each time step. To initialize the fit, $n_0$ is set to the location of the maximum, $c$ to the approximate width of 8 positions, which is equivalent to the spatial width of the wave packet in the first time step, and $A$ is set to the maximum intensity per time step. By extracting the error margins of the fit parameter $n_0$ for a confidence interval of 95% it is possible to estimate the uncertainty of this procedure and so to calculate $\Delta n_j^{\pm \varphi_0}$: the uncertainty of the fit for the positive and negative phase modulation. The total uncertainty in the integrated Berry curvature due to the fit is thus given by $\sqrt{\left(\Delta n_j^{+\varphi_0}\right)^2 + \left(\Delta n_j^{-\varphi_0}\right)^2}$, while the uncertainty with respect to $Q$ is estimated by the spectral width $\Delta Q$ of the wave packet, as determined by numerically Fourier transforming the initial distribution. Here, a value of approximately $\Delta Q \approx 0.05\pi$ is found for the FWHM of the intensity for all data sets.

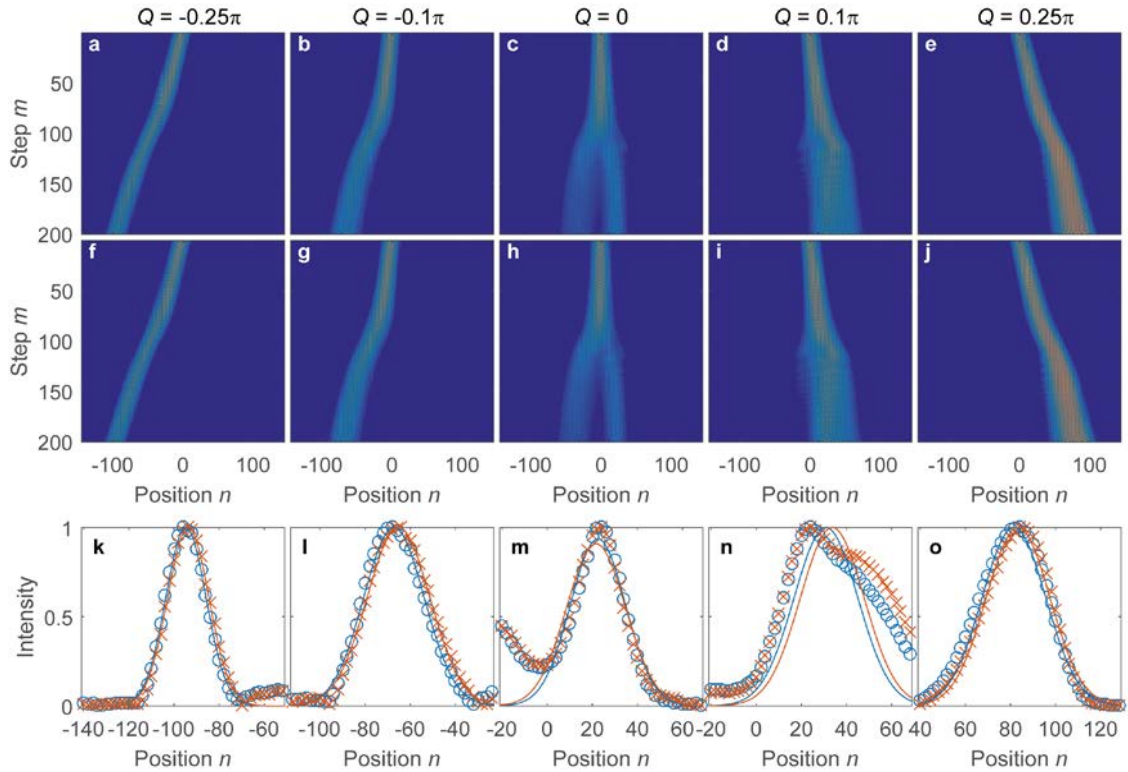

**Supplementary Figure S8 | Experimental propagation for the two-step modulation for the upper band.** Measurement of a wave packet in the upper band for $-\varphi_0$ (**a-e**) and for $+\varphi_0$ (**f-j**) over 200 time steps. At each time step, the center-of-mass is estimated by applying a Gaussian fit to the cross section. (**k-o**) As an example of this procedure, the cross sections and fitted curves are displayed at the last time step for $-\varphi_0$ (red) and for $+\varphi_0$ (blue).

By using this protocol, the integrated Berry curvature is directly observed. However, to reconstruct the Berry curvature itself, it is necessary to differentiate the experimental results numerically. Since this procedure is very sensitive to noise, the experimental data is first smoothed by an averaging filter spanned over five elements. The influence of the averaging filter is shown in Supplementary Figure S9.

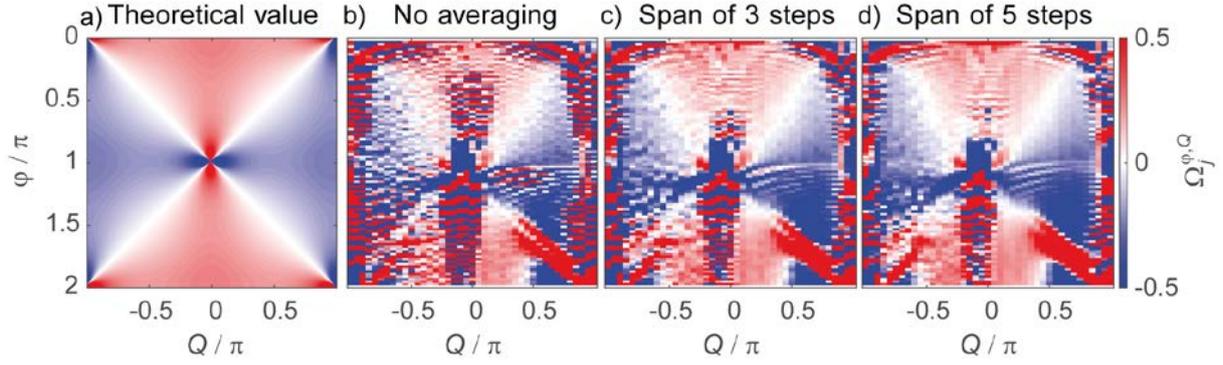

**Supplementary Figure S9 | Influence of the averaging on the reconstruction of the Berry curvature.** High frequency noise due to experimental imperfections is reduced by an averaging filter. In Figure 3 of the manuscript, the filter spans 5 data points (d) which effectively suppresses the noise that it is inevitably present in the raw data (b).

## S7 Spectral width of the wave packet

All experiments which are discussed in the manuscript are carried out with a wave packet covering approximately 8 positions in real space corresponding to a spectral width of $0.05\pi$. In our semi-classical analysis, we assume that the wave packet propagates like a classical particle with a well-defined group velocity that is shifted by the anomalous velocity due to the Berry curvature. However, this description is no longer applicable if the wave packet becomes too spectrally broad in momentum, as then it is no longer sufficient to characterize the motion of the wave packet by one value of the Bloch momentum $Q$. In order to investigate the dependence of our results on the width of the wave packet, the measurement has been reproduced in simulations for wave packets of different width (see Supplementary Fig. S10).

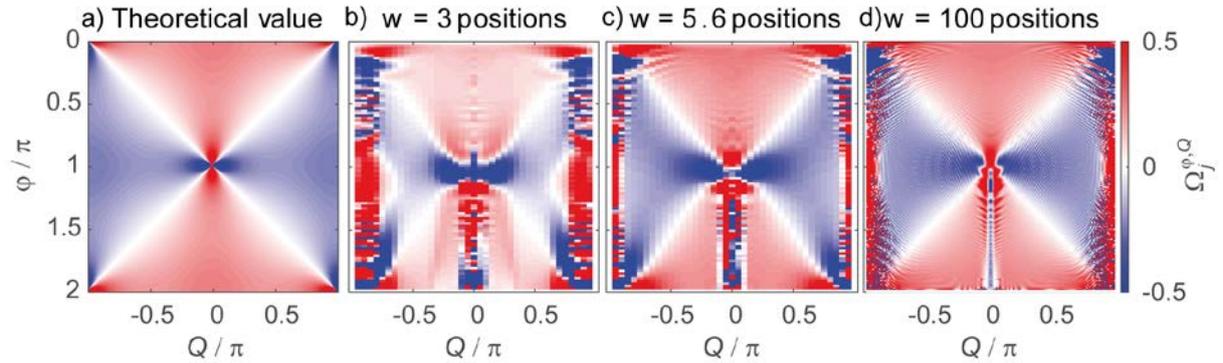

**Supplementary Figure S10 | Dependence on the spectral width of the wave packet (simulation).** While the basic characteristics of the Berry curvature **(a)** are already recovered by a wave packet with a width, $w$, of only 3 positions in real space **(b)**, the resolution is sharpened with growing width of the wave packet, as this corresponds to a narrower spread in Bloch momenta. In the experiment, similar parameters to those (c) are used; the width here of 8 positions in real space corresponds to a spectral width of $0.05\pi$, with the Bloch momentum scanned in steps of $0.05\pi$. For (b,c) the same gradient $\varphi_0 = \pi/50$ is used, while in (d) a smaller gradient $\varphi_0 = \pi/100$ and a finer discretization of $Q$ in steps of $0.01\pi$ is chosen. Additionally, in (d) the averaging filter spans 19 time steps.

## S8 Anomalous transport for the four time step modulation

In order to demonstrate the general applicability of the experimental protocol, a second modulation scheme is discussed in this section. It is designed in the same way as the previous modulation and differs only in the internal length of the modulation scheme. For the second modulation

$$\Phi_{4\text{step}}(M(m)) = \begin{cases} -\varphi, & \mod(m, 4) = 0 \text{ or } 1 \\ +\varphi, & \mod(m, 4) = \text{else} \end{cases}, \quad (S16)$$

where now the modulation period $M(m) = \lfloor m/4 \rfloor$ lasts for four time steps. Therefore, the dispersion relation is calculated for one modulation period using the ansatz

$$\begin{pmatrix} u_n^m \\ v_n^m \end{pmatrix} = \begin{pmatrix} U \\ V \end{pmatrix} e^{-\frac{i\theta m}{4}} e^{\frac{iQn}{4}}. \quad (S17)$$

In this case, the dispersion relation reads as

$$\cos\theta = \frac{1}{4}(\cos 2Q - \cos 2\varphi - 4\cos Q \cos\varphi). \quad (S18)$$

Interestingly for $\varphi = 0$, when no modulation is present, the size unit cell collapses again to two time steps, hence the band structure represented by Eq. (S18) is artificially back. This aspect is further illustrated by comparing the band structure for $\varphi = 0$ given by Eq. (S3) and Eq. (S18). As one can see, for the four time step ansatz, this back-folding leads to a closing of the band gap, which is not physically-motivated but instead an artefact of the chosen ansatz in Eq. (S17). However, for any finite phase modulation $\varphi \neq 0$, the four time step ansatz is needed in order to cover the whole modulation period. The dispersion relation for the four-step modulation is shown in Supplementary Fig. S11a and b for various values of $\varphi$.

By applying the procedure for the two-step modulation also for the four-step scheme, the results presented in Supplementary Fig. S11 are obtained. The back-folding of the band structure along $\theta$ leads to additional singularities in the case of the Berry curvature (see Supplementary Fig. S11a-c). For the propagation in real space, the touching of the bands leads to an emission of a smaller counter-propagating secondary wave; to account for this, we fit the propagation cross-section with a double Gaussian profile, instead of a single Gaussian as in Eq. (S15), taking the center of the higher Gaussian peak as the center-of-mass position. In this way, we can again extract the integrated Berry curvature, as shown in Supplementary Fig. S11d. In a final step, by numerically differentiating the motion of the center-of-masses the Berry curvature for the four step modulation is recovered (see Supplementary Fig. S11e).

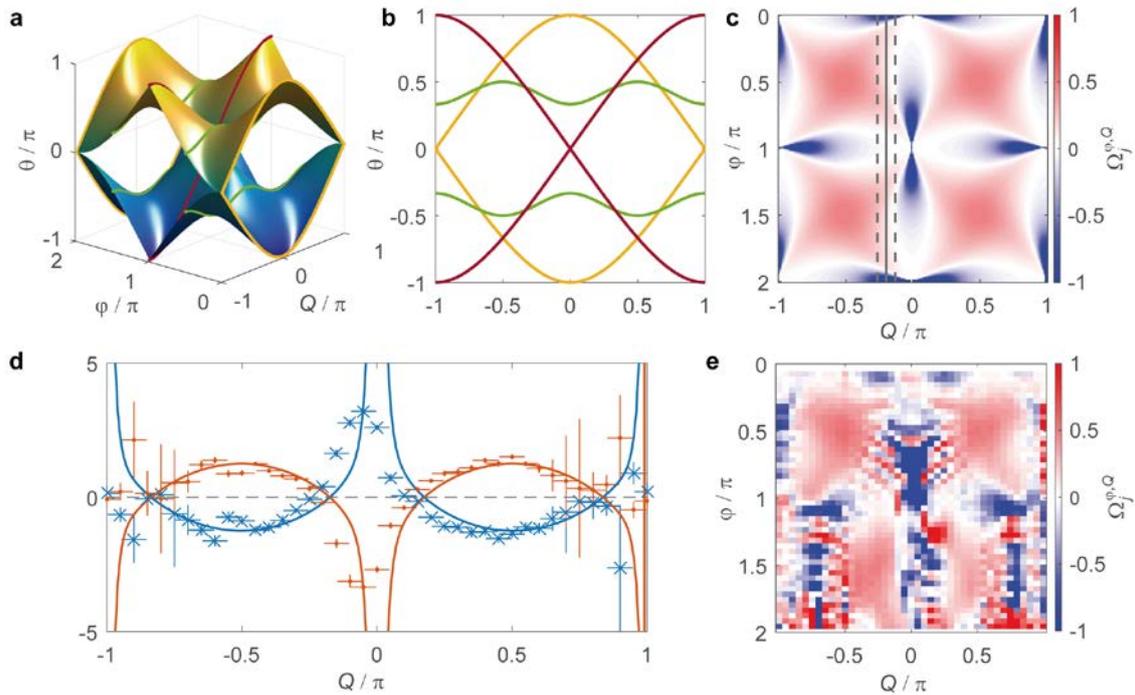

**Supplementary Figure S11| Anomalous transport for the four-step modulation. a,** Band structure of the system. As for the two-step modulation, the band structure is also periodic in $\theta$. **b,** Cut through of the dispersion relation for $\varphi = 0$ (yellow), $\varphi = \pi/2$ (green) and $\varphi = \pi$ (red). Compared to the two-step modulation, additional band-degeneracy points are present, as also visible as singularities in the theoretical plot of the Berry curvature for the lower band **(c)**. **d,** Measured integrated Berry curvature for the four-step modulation for the lower band (orange dots) and the upper (blue crosses). Again, close to the singularities at $Q = 0, \varphi = \pm\pi$, the protocol fails as adiabaticity breaks down. **e**, Reconstruction of the Berry curvature of the lower band based on the four step modulation. The results are in a good qualitative agreement to the theoretical value in *(c)*.

# Supplementary References